\documentclass[12pt]{article}

\RequirePackage[OT1]{fontenc}
\usepackage{amsthm,amsmath,natbib}
\usepackage{array}
\usepackage{float}
\usepackage{booktabs}
\usepackage{tabularx}
\usepackage{xcolor}
\usepackage{color}
\RequirePackage[colorlinks,citecolor=blue,urlcolor=blue]{hyperref}
\usepackage{alltt}
\usepackage{fancyvrb}
\usepackage{algorithm2e}
\usepackage{pdfpages}
\usepackage{multirow}
\usepackage{lineno}
\usepackage{comment}
\usepackage{verbatim}
\usepackage{lscape}
\usepackage{graphicx}
\usepackage{eurosym}
\usepackage{gensymb}
\usepackage{graphicx}

\def\bi{\begin{itemize}}
\def\ei{\end{itemize}}
\def\be{\begin{equation}}
\def\ee{\end{equation}}

\begin{document}
\thispagestyle{empty}
\baselineskip=27pt
\vskip 4mm
\begin{center} 

{\Large{\bf From scenario-based seismic hazard to scenario-based landslide hazard: rewinding to the past via statistical simulations}}

\end{center}

\baselineskip=12pt
\vskip 3mm

\begin{center}
\large
Luguang Luo$^{1}$, Luigi Lombardo$^{2*}$, Cees van Westen$^{2}$, Xiangjun Pei$^{1}$, Runqiu Huang$^{1}$ 
\end{center} 

\footnotetext[1]{
\baselineskip=10pt State Key Laboratory of Geohazard Prevention and Geoenvironment Protection (SKLGP), Chengdu University of Technology, Sichuan, 610059, China}

\footnotetext[2]{
\baselineskip=10pt University of Twente, Faculty of Geo-Information Science and Earth Observation (ITC), PO Box 217, Enschede, AE 7500, Netherlands}

\baselineskip=16pt

\begin{center}
{\large{\bf Abstract}}
\end{center}

The vast majority of landslide susceptibility studies assumes the slope instability process to be time-invariant under the definition that ``the past and present are keys to the future". This assumption may generally be valid. 
However, the trigger, be it a rainfall or an earthquake event, clearly varies over time. And yet, the temporal component of the trigger is rarely included in landslide susceptibility studies and only confined to hazard assessment. 

In this work, we investigate a population of landslides triggered in response to the 2017 Jiuzhaigou earthquake ($M_w = 6.5$) including the associated ground motion in the analyses, these being carried out at the Slope Unit (SU) level. We do this by implementing a Bayesian version of a Generalized Additive Model and assuming that the slope instability across the SUs in the study area behaves according to a Bernoulli probability distribution. 

This procedure would generally produce a susceptibility map reflecting the spatial pattern of the specific trigger and therefore of limited use for land use planning. However, we implement this first analytical step to reliably estimate the ground motion effect, and its distribution, on unstable SUs. 

We then assume the effect of the ground motion to be time-invariant, enabling statistical simulations for any ground motion scenario that occurred in the area from 1933 to 2017. 

As a result, we obtain the full spectrum of potential susceptibility patterns over the last century and compress this information into a susceptibility model/map representative of all the possible ground motion patterns since 1933. This backward statistical simulations can also be further exploited in the opposite direction where, by accounting for scenario-based ground motion, one can also use it in a forward direction to estimate future unstable slopes.

\baselineskip=10pt

\par\vfill\noindent

\newpage
\baselineskip=16pt

\section{Introduction}
``The past and present are keys to the future" has been the underlying principle over three decades to support landslide susceptibility studies \citep[e.g.,][]{calvello2013,ercanoglu2004,Ermini2005,varnes1984}. This hypothesis implies time-invariance of the slope response. However, if on the one hand it may be true that the effect of predisposing factors and triggers does not change over time because the law of physics stay the same; it is certainly true that the space-time patterns of the triggers change from one event to another \citep{westen2008}. And yet, the temporal dimension is rarely accounted for \citep{corominas2008} despite the susceptibility of a given area should change as a function of the space-time realization of the trigger \citep{ghosh2012,lombardo2019}.  

To deal with such complexity, the research community dealing with data-driven landslide susceptibility assessment typically follows two lines of analysis: 

\begin{itemize}
    \item the signal of the trigger is ignored in the landslide susceptibility models \citep[e.g.,][]{cama2015predicting,reichenbach-2018}. 
    This procedure results in predictive maps of landslide occurrence for a given area, ignoring the specific impacts produced by a given trigger. The strength of this procedure consists of delivering simple realizations of the geomorphological responses. And, in practice this is often the only possibility, when the available landslide inventory data lacks the information on date of occurrence \citep{guzzetti2012}. However, the main downside is due to the fact that the spatial \citep{Lombardo2018b} and temporal \citep{Samia2017} dependence of the landslide distribution is entirely neglected. More specifically, the spatial distribution of the trigger intensity induces some degree of dependence in the landslide distribution \citep{lombardo2019geostatistical} and the resulting landslides further induce some degree of temporal dependence to subsequent occurrences because of local disturbance, re-activation and re-mobilization processes \citep{samia2017b}.  
    
    \item the signal of the trigger is incorporated in the predictive models. This is generally done in near-real time applications where the space-time dynamics of the trigger are added to a baseline susceptibility. Such examples can be found in \citet{Kirschbaum2010,kirschbaum2018satellite} in case of rainfall-induced landslides and in \citet{nowicki2018global,tanyas2019global} for the earthquake counterpart. The strength of this type of applications resides in a much closer realization of the hazard definition \citep{varnes1984,fell2008guidelines,Guzzetti1999}, where the probability of occurrence is estimated in a given time period and area.   
\end{itemize}
       
The latter case is inevitably much more comprehensive than the former, but it comes with an added level of complexity, and at times even non-feasibility because of the lack of information on the trigger. In fact, an initial stage is required where a susceptibility model is calibrated by using an event-based landslide inventory together with traditional morphometric and thematic properties, as well as the actual pattern of the trigger. On the basis of the calibration stage, the regression coefficients associated with the trigger are estimated and used in a second phase to make a prediction. The prediction is essentially a constantly changing susceptibility/hazard model, where the change is driven by substituting the trigger with near-real time estimates coming from remotely-sensed precipitation, for rainfall-induced landslides \citep{kirschbaum2018satellite}, or by ground motion parameters for seismically induced landslides \citep{nowicki2018global}. 

Currently, these procedures have been implemented by keeping the regression coefficient of the trigger fixed. In other words, the uncertainty around the estimation of this model component has been neglected. 
In this work, we take a very similar starting point, but accounting for the uncertainty of the trigger effect via statistical simulations. 
More specifically, by definition any Bayesian model returns the distribution of each model component. By taking advantage of this structure, we calibrate and validate an initial model where the earthquake-induced landslides caused by the 2017 Jiuzhaigou earthquake ($M_w = 6.5$) are modeled via a Bayesian Generalized Additive Model (GAM) by incorporating conditioning factors together with a triggering factor expressed in the form of the Peak Ground Acceleration (PGA) of the Jiuzhaigou earthquake. As a result, we extract the full distribution of the regression coefficient associated with the PGA and simulate thousands realizations of the susceptibility by substituting the trigger pattern with the analogous ground shaking parameters belonging to any past earthquake, for which we have accessible records, for a time window between 1933 and 2017.

Under the assumption that the functional relationship between the trigger and the landslides is well estimated, and that other causative factors have not significantly changed through time, this procedure allows one to reconstruct the space-time variation of the susceptibility under different environmental stresses and to retrieve the distribution of unstable slope throughout the investigated time window, for a given study area.  

The manuscript is arranged as follows: in Section \ref{sec:StudyArea}, we describe the study area, the characteristics of the inventory associated with the Jiuzhaigou earthquake as the data on the previous earthquakes in the period since 1933. In Section \ref{sec:Strategy} we describe the subdivision of the area in mapping units, their status (landslide/no-landslide assignment), the explanatory variables' selection, the modeling framework and the simulations' structure. Section \ref{sec:Results} presents the results which will be discussed and interpreted in Section \ref{sec:Discussion}. Section \ref{sec:Conclusions} summarizes the take home message of this work and the implications of what we propose.

\section{Study Area}\label{sec:StudyArea}

\subsection{Geomorphological settings}\label{sec:Settings}

\begin{figure}[t!] 
	\centering
	\includegraphics[width=\linewidth]{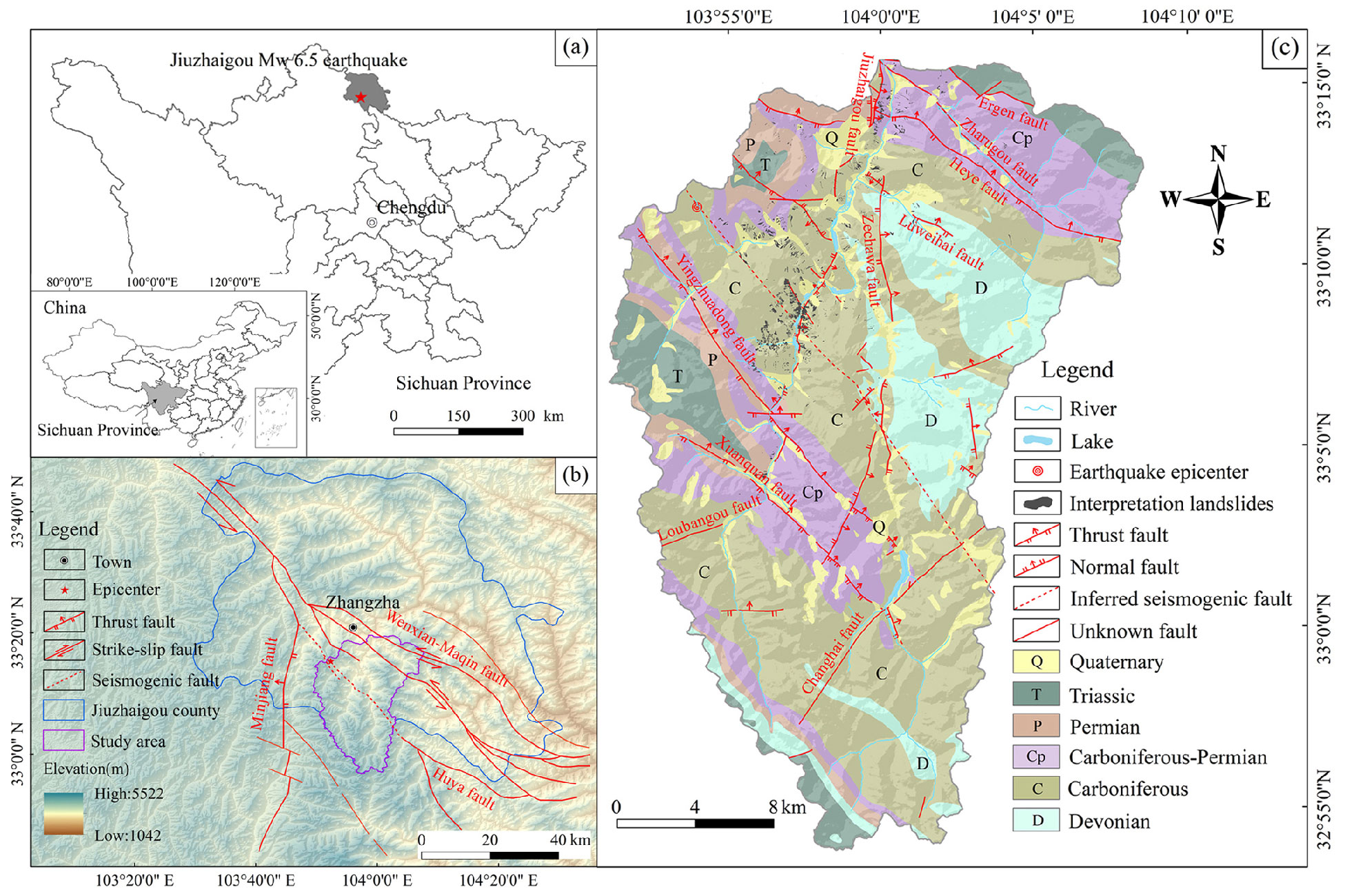}
	\caption{(a) Geographic context; (b) study area and large scale tectonic setting; (c) Geological map, showing the small scale tectonic setting as well as the Jiuzhaigou-earthquake-induced landslide inventory.}
	\label{Figure1}
\end{figure}

The study area, known as Jiuzhaigou National Geopark, is located in the Jiuzhaigou County, near the northern boundary of Sichuan province in the southwest of China (see Fig.\ref{Figure1}a). It is a part of the Min Mountains range between the Tibetan Plateau and Sichuan Basin, approximately $285 km$ north of the city of Chengdu. Jiuzhaigou was recognised as a World Heritage Site by UNESCO in 1992 and a World Biosphere Reserve in 1997. It is one of the most popular tourist destinations in the region and more than five million tourists visit this outstanding natural landscape each year. Jiuzhaigou is the main tributary of Baishui River, and is one of the sources of Jialing River via the Bailong River, part of the Yangtze River system. 

The area features high-altitude karst landscape shaped by glacial, hydrological and tectonic activity (Fig.\ref{Figure1}b). The latter results from the influence of the Minjiang fault (northwestern section), the Wenxian-Maqin fault and the Huya fault \citep{Yi2018}. The Huya fault is dominated by left-lateral strike slip. Previous studies have pointed out the north-west section of the Huya fault to be the specific seismogenic source of the Jiuzhaigou earthquake \citep{fan2018,Li2017,wu2018}. More generally, the Jiuzhaigou area is a seismically active hilly mountainous region which is subjected to more than 50 events with $M\geq5$ in the past century \citep{fan2018}. 

The study area extends from latitudes 32$^{\circ}$54’21”N to 33$^{\circ}$16’9”N and from longitudes 103$^{\circ}$46’24”E to 104$^{\circ}$3’54”E, covering an extent of approximately $653 km^2$ (Figure \ref{Figure1}c). 
This region belongs to a typical cold semi-arid monsoon climate with annual precipitation of about $704.3mm$. The mean annual temperature is 7.8$^{\circ}C$, with a minimum of -3.7$^{\circ}C$ in January and maximum of 16.8$^{\circ}C$ in July.

Topographically, the elevation ranges from $2000 m$ to $4828 masl$. The study area encompasses three valleys namely, Shuzheng, Rize and Zechawa valleys, arranged in a Y shape. The slope gradients, derived from a Digital Elevation Model with a spatial resolution of $30 m$ range from $0$ to $78\degree$ being generally higher than $30\degree$ on average. The main lithological formations in the study area are Devonian, Carboniferous, Permian, Triassic, Dolomite and Quaternary and consist of carbonate rocks such as dolomite and tufa, as well as some sandstone and limestone (see Figure \ref{Figure1} and Table \ref{table:LitoTab} in appendix). Because of the complex tectonic settings, ten main and several small scale faults dissect the carbonatic lithotypes, leaving the rock masses weakened by a large number of joints and cracks.

\subsection{Jiuzhaigou earthquake}\label{sec:Jiuzhaigou}

On $8^{th}$ August 2017, an earthquake of magnitude $M_w~6.5$ struck the Jiuzhaigou county, belonging to the Sichuan province, China. It was the third strong earthquake in the region in the past 11 years, after the 2008 $M_w$ 7.9 Wenchuan earthquake and the 2013 $M_w$ 6.6 Lushan earthquake \citep{fan2018}. The epicentre of this earthquake was only $5 km$ west of the Jiuzhaigou National Park, where the touristic infrastructure of the UNESCO world heritage site was damaged, 31 person were killed and 525 were injured \citep{wang2018}. 
This event affected to different extents more than $175,000$ people, both tourists and locals. More than $73,000$ buildings were damaged $76$ of which collapsed. The scenic spot was temporarily closed after the earthquake and reopened only after two years (on October 8, 2019), which severely affected the economy around Jiuzhaigou and the Aba Prefecture.

The ground motion not only directly damaged properties and lives but, in a cascading effect, it also caused widespread landsliding, which in turn contributed to increase the overall losses. 

\subsection{Landslide mapping}\label{sec:Landslides}

The preparation of a reliable and accurate landslide inventory map recording the location, spatial extent and landslide characteristics is essential for any susceptibility analysis \citep{guzzetti2012}. In case of earthquake-induced landslides (EQIL), the quality and completeness of an inventory affects the susceptibility estimates of any landslide affected area \citep{tanyas2018}. 

For this reason, we undertook a multi-source mapping procedure to discriminate pre- and co- seismic landslides. The landslide inventory was carried out through visual interpretation of high-resolution images with different resolutions and sources (see Table \ref{table:DataSum}). Remotely sensed scenes, Unmanned Aerial Vehicle (UAV) photographs and subsequent detailed field surveys were used for mapping the landslides source and deposition areas.

\begin{figure}[t!] 
	\centering
	\includegraphics[width=\linewidth]{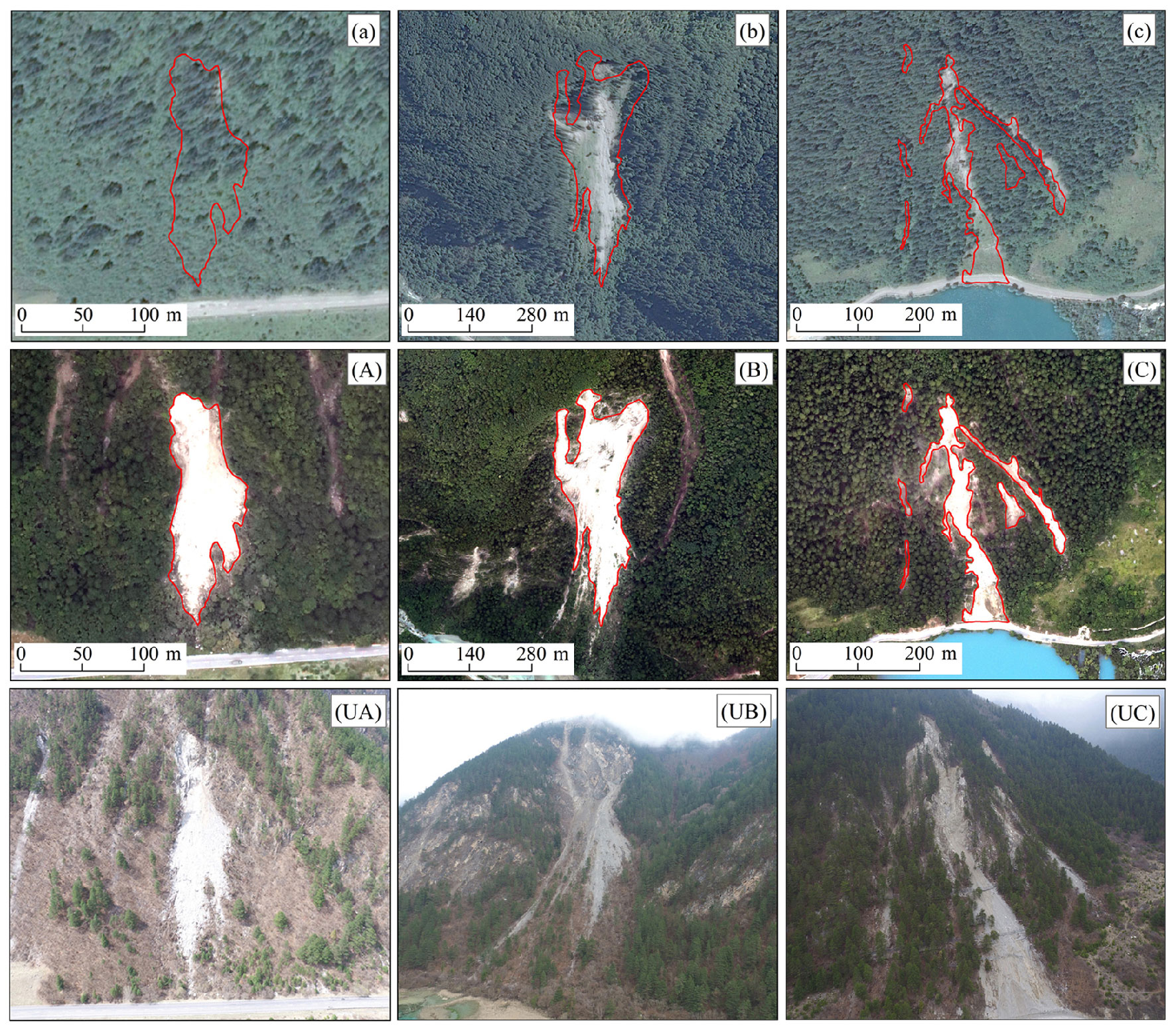}
	\caption{Examples of landslide interpretation based on pre- and post-earthquake high resolution satellite and UAV images. Panels a, b and c correspond to pre-earthquake conditions; panels A, B, C shows post-earthquake satellite images; panels UA, UB, UC are photographs acquired using drones.}
	\label{Figure2}
\end{figure}

More specifically, the existence of landslides prior to the earthquake was investigated by using a Spot-5 scene ($2.5 m$ resolution) acquired on 21 December 2015. These pre-earthquake inventories were used to isolate co-seismic landslides (new activations and re-activations) mapped via Gaofen-1 and Gaofen-2 satellite images ($1 m$ resolution) acquired on 16 August and 9 August 2017, respectively. And, we also refined the mapping procedure by additionally examining UAV photos ($0.2 m$ resolution) of key areas acquired during field surveys (Table \ref{table:DataSum}). Some examples of identified landslides are shown in Figure \ref{Figure2}.

\begin{table}[t!]
\centering
\begin{tabular}{|c|c|c|c|c|}
\hline
\textbf{Data description} & \textbf{Data Type} & \textbf{Resolution} & \textbf{Source}\\ 
\hline
Pre-EQ satellite image & Raster (*.tif) & $2.5 m$ & Spot-5\\
\hline
Post-EQ satellite image & Raster (*.tif) & $1.0 m$ & Gaofen-1 and Gaofen-2\\
\hline
Post-EQ UAV aerial image & Raster (*.tif) & $0.2 m$ & Field survey\\
\hline
\end{tabular}
\caption{Overview of the data used for mapping landslides. EQ stands for earthquake.} 
\label{table:DataSum}
\end{table}

Overall $1022$ landslides have been associated with the Jiuzhaigou earthquake. The total landslide area sums up to $3.88 km^2$ covering approximately 0.6\% of the study area. They mainly occurred along the valleys, roads or rivers (see Figure \ref{Figure2}). 

\begin{figure}[t!] 
	\centering
	\includegraphics[width=0.6\linewidth]{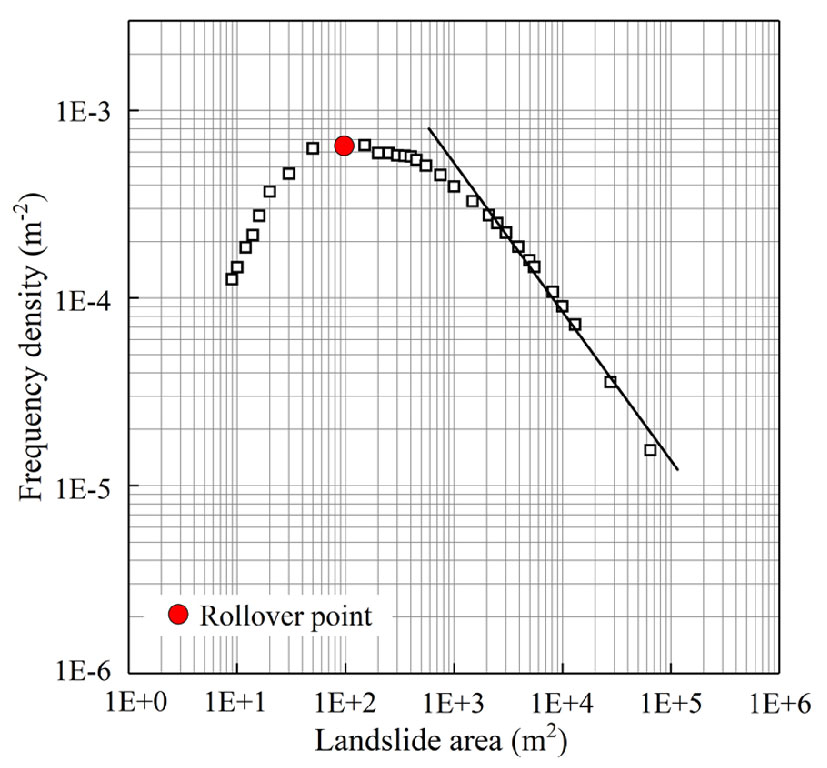}
	\caption{Frequency Area Distribution of co-seismic landslides.}
	\label{Figure3}
\end{figure}

The failure mechanisms consisted of shallow rock or debris slides with minor rockfall occurrences \citep{varnes1978,Hungr2014} spanning from small to moderate landslide in size. This is shown in Figure \ref{Figure3} where the Frequency Area Distribution shows a rollover point at approximately $100 m^2$, a minimum landslide area of around $30 m^2$ and a well recognizable distribution, traditionally explained by a Inverse-Gamma distribution \citep{fan2019}. These characteristics have been recently recognized for earthquake-induced landslide inventories of good quality and completeness \citep{tanyas2018,Tanyas2020}. 

\subsection{Earthquake history}\label{sec:Quakes}

The Jiuzhaigou National Geopark is located in the transition zone of the western margin of the Sichuan Basin and the Qinghai-Tibet Plateau which features tectonically active mountains characterized by narrow and steep valleys. Numerous moderate to large earthquakes have been recorded in the past century, with known record of associated landslides. Specifically, spanning a $200km$ radius from the Jiuzhaigou epicenter, the USGS earthquake catalog \citep{earle2009} reports 76 earthquakes with magnitude between 5 and 8. Among them, seven earthquakes have a magnitude $6.0\leq M_W \leq7.0$, and only one above 7.0, which corresponds to the Diexi earthquake. 

\begin{figure}[t!] 
	\centering
	\includegraphics[width=\linewidth]{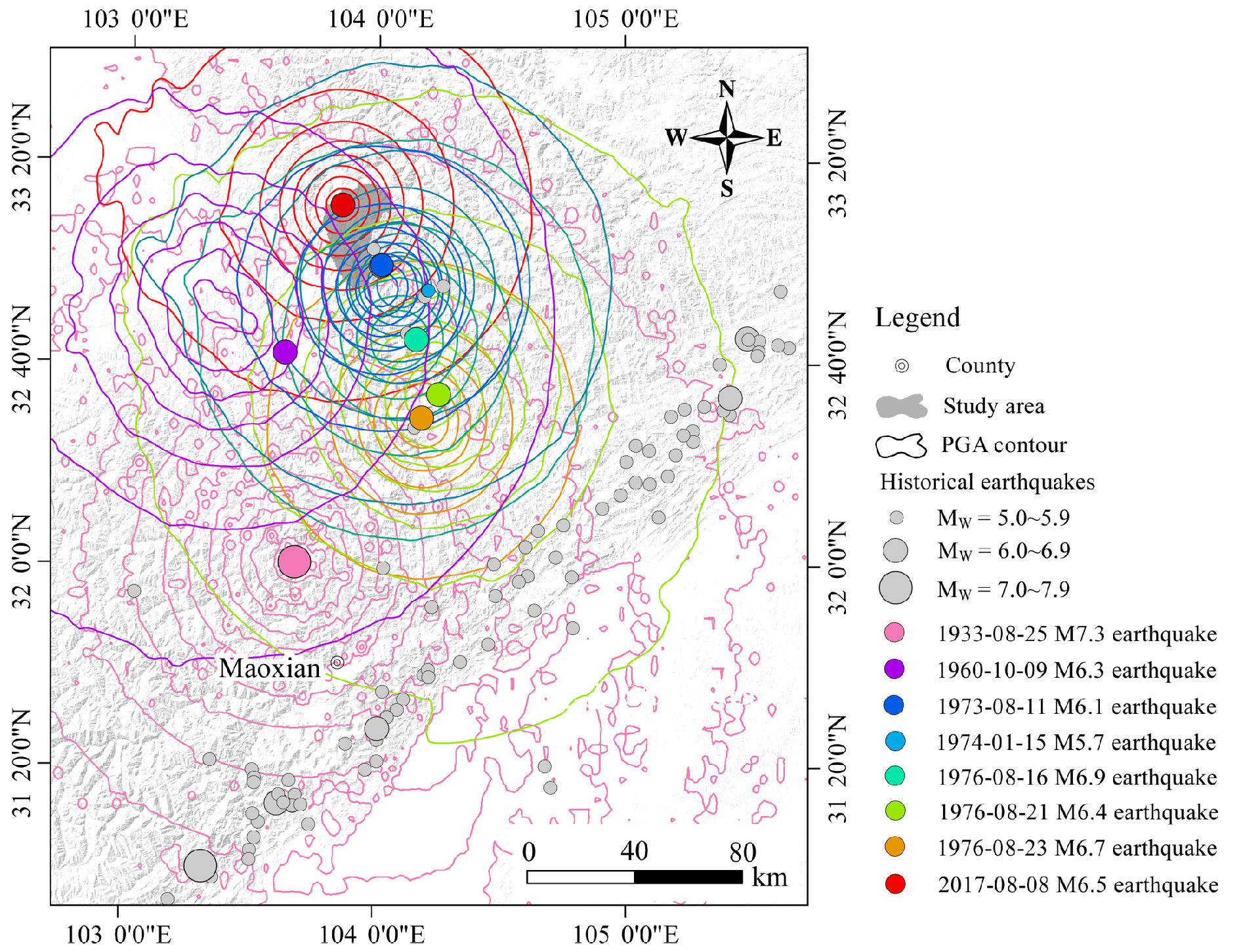}
	\caption{Location of the epicenters and associated PGA levels (shown as contour lines) of all the earthquakes available in the  ShakeMap Atlas \citep{allen2008}, which have struck the area since 1933.)}
	\label{Figure4}
\end{figure}

To reconstruct the susceptibility patterns due to past earthquakes, we examined all the available earthquakes which have affected the study area and for which the associated ground motion characteristics were available at the ShakeMap service of the US Geological Survey (USGS) \citep{allen2008,wald2008}. We collated 7 past earthquakes here listed in Table \ref{table:QuakesList} and geographically shown in Fig.\ref{Figure4}.
The $M_W$ range since 1933 spans from a minimum of 5.7 to a maximum of 7.3, with most of the epicenters located to the south and outside the boundary of the study area. And, two more recent earthquakes which occurred to the north and within the study area.

\begin{table}[t!]
\centering
\scalebox{0.79}{
\begin{tabular}{|c|c|c|c|c|c|c|}
\hline
\textbf{ID} & \textbf{Location} & \textbf{Date/time} & \textbf{Epicentre Lat} & \textbf{Epicentre Long} & \textbf{$M_w$} & \textbf{Depth (km)}\\  
\hline
a & Diexi & 1933-08-25/07:50:33 (UTC) & 32.012\degree N & 103.676\degree E & 7.3 & 15 \\
\hline
b & Songpan & 1960-10-09/10:43:45 (UTC) & 32.706\degree N & 103.629\degree E & 6.3 & 25 \\
\hline
c & Songpan & 1973-08-11/07:15:39 (UTC) & 32.995\degree N & 104.015\degree E & 6.1 & 33 \\
\hline
d & Songpan & 1974-01-15/22:50:29 (UTC) & 32.913\degree N & 104.203\degree E & 5.7 & 33 \\
\hline
e & Songpan-pingwu & 1976-08-16/14:06:45 (UTC) & 32.752\degree N & 104.157\degree E & 6.9 & 16 \\
\hline
f & Songpan-pingwu & 1976-08-21/21:49:54 (UTC) & 32.571\degree N & 104.249\degree E & 6.4 & 33 \\
\hline
g & Songpan-pingwu & 1976-08-23/03:30:07 (UTC) & 32.492\degree N & 104.181\degree E & 6.7 & 33 \\
\hline
h & Jiuzhaigou & 2017-08-08/13:19:49 (UTC) & 33.193\degree N & 103.855\degree E & 6.5 & 9 \\
\hline
\end{tabular}}
\caption{List of eartquake which have affected the study area and for which the shaking levels were digitally accessible in the ShakeMap Atlas \citep{allen2008}.} 
\label{table:QuakesList}
\end{table}

\begin{figure}[t!] 
	\centering
	\includegraphics[width=\linewidth]{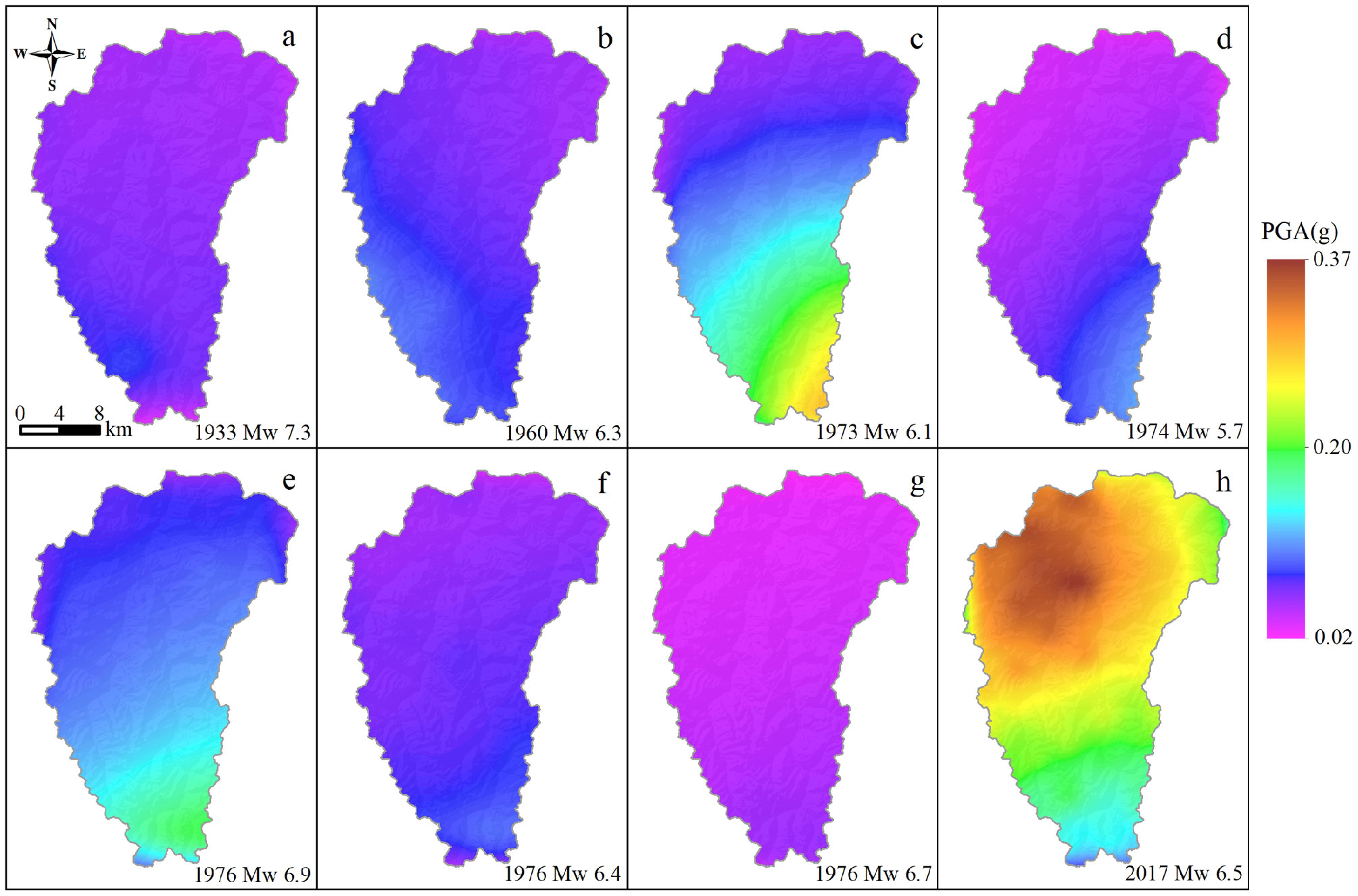}
	\caption{PGA patterns affecting the study area for each of the earthquakes under examination \citep[Source: USGS ShakeMap system][]{garcia2012}. Notably, there is no strict minimum PGA reported in the literature to trigger landslides \citep{sassa2007}. However, several articles have reported that the vast majority of earthquake-induced landslides trigger with a minimum threshold of $0.05g$ \citep{jibson2016,Tanyas2019}, which is also contained in every PGA map shown in this figure: (a) $0.03<PGA< 0.07$; (b) $0.03<PGA<0.1$; (c) $0.04<PGA<0.25$; (d) $0.02<PGA<0.1$; (e) $0.04<PGA<0.2$; (f) $0.04<PGA<0.09$; (g) $0.02<PGA<0.05$; (h) $0.08<PGA<0.37$;}
	\label{Figure5}
\end{figure}

This is more evident in Figure \ref{Figure5} where we focus on the study area. In fact the highest PGAs are recorded in the northern sector, these being associatd with the 2017 Jiuzhaigou earthquake. The remaining patterns generally show a northward increase of PGA levels.   

\section{Modeling strategy}\label{sec:Strategy}

\subsection{Mapping unit}\label{sec:SU}

The first requirement of any landslide susceptibility model is the choice of the type of mapping unit used in the statistical analysis. The most common choice corresponds to a regular squared lattice or grid \citep[e.g.,][]{cama2016,Hussin2016,reichenbach-2018}. However, this mapping unit is sensible to mapping errors \citep{steger2016propagation} and the assignment of the instability status is inherently uncertain as it is often subjectively chosen between the centroid of the landslide body \citep[e.g.,][]{Hussin2016,Camilo2017} or the highest point along the landslide polygon \citep[e.g.,][]{amato2019,Lombardo2014}. 

To avoid the issues concerning the grid-based choice, here we opted for a Slope Unit (SU) partition \citep{carrara1995}. We computed the SUs by using the \emph{r.slopeunits} software made accessible by \citet{alvioli2016}. More specifically, we parameterized \emph{r.slopeunits} with a very large Flow Accumulation threshold ($800,000 m^2$) to scan the area with a large configuration of possible SU arrangements, whose conversion to the final setup we controlled via a small minimum slope unit area of $10,000 m^2$.

When at least one centroid of a landslide initiation polygon was located in a SU, it was assigned a positive landslide status, and the others were assigned a non-landslide status. 

\subsection{Covariates}\label{sec:Covar}

The covariate set we chose features eight morphometric properties, two related to the geological setting, and one related to the Peak Ground Acceleration (Table \ref{table:CovarList}). 

The topographic covariates were derived from the a $30 m$ resolution DEM provided by Sichuan Bureau of Surveying, Mapping and Geoinformation, using a number of different methods, following the references indicated in Table \ref{table:CovarList}. 

The same institute provided the Geological Map of the area, which we rasterized to coincide with the DEM resolution. We also generated a bedding map by digitizing strike and dip measurements collected in the field and subsequently crossing this parameters together with the exposition of any given lithotype reported in the geological map (see Table \ref{table:BeddingTab} in Appendix A). More specifically, we followed the idea introduced by \citet{ghosh2010} and the approach later proposed by \citet{santangelo2015} exploiting a total sample of 1490 dip measurement, which we then grouped by lithology, as follows:
125 in Q, 123 in T, 133 in P, 368 in Cp, 560 in C, 79 in Dc, 102 in Dd. 

From the fault map provided by the Exploitation of Mineral Resources, we computed the Euclidean distance from each pixel in the study area to the nearest faultline line. And, we repeated the same operation to compute the Euclidean distance from the river network. 

Because the SUs actually contain a distribution of pixel values for each covariate, we summarized the covariate information at the SU level by computing mean ($\mu$, hereafter) and standard deviation ($\sigma$, hereafter) of continuous properties. As for the categorical coveriates such as bedding and geology, we extracted the most represented class per SU.   

\begin{table}[t!]
\centering
\scalebox{0.9}{
\begin{tabular}{|c|c|c|c|}
\hline
\textbf{Covariate name} & \textbf{Scale/Resolution} & \textbf{Reference} & \textbf{Acronym} \\  
\hline
\multirow{2}{*}{Elevation} & \multirow{2}{*}{$30 m$} & \multirow{2}{*}{/} & $Elev_\mu$ \\ \cline{4-4}
                           &                         &                     & $Elev_\sigma$ \\
\hline
\multirow{2}{*}{Slope} & \multirow{2}{*}{$30 m$} & \multirow{2}{*}{\citet{zevenbergen1987}} & $Slope_\mu$ \\ \cline{4-4}
                           &                         &                     & $Slope_\sigma$ \\
\hline
\multirow{2}{*}{Profile Curvature} & \multirow{2}{*}{$30 m$} & \multirow{2}{*}{\citet{heerdegen1982}} & $PRC_\mu$ \\ \cline{4-4}
                           &                         &                     & $PRC_\sigma$ \\
\hline
\multirow{2}{*}{Planar Curvature} & \multirow{2}{*}{$30 m$} & \multirow{2}{*}{\citet{heerdegen1982}} & $PRC_\mu$ \\ \cline{4-4}
                           &                         &                     & $PLC_\sigma$ \\
\hline
\multirow{2}{*}{Eastness} & \multirow{2}{*}{$30 m$} & \multirow{2}{*}{\citet{Lombardo2018c}} & $EN_\mu$ \\ \cline{4-4}
                           &                         &                     & $EN_\sigma$ \\
\hline
\multirow{2}{*}{Northness} & \multirow{2}{*}{$30 m$} & \multirow{2}{*}{\citet{Lombardo2018c}} & $NN_\mu$ \\ \cline{4-4}
                           &                         &                     & $NN_\sigma$ \\
\hline
\multirow{2}{*}{Relative Slope Position} & \multirow{2}{*}{$30 m$} & \multirow{2}{*}{\citet{bohner2006}} & $RSP_\mu$ \\ \cline{4-4}
                           &                         &                     & $RSP_\sigma$ \\
\hline
\multirow{2}{*}{Topographic Wetness Index} & \multirow{2}{*}{$30 m$} & \multirow{2}{*}{\citet{beven1979}} & $TWI_\mu$ \\ \cline{4-4}
                           &                         &                     & $TWI_\sigma$ \\
\hline
\multirow{2}{*}{Peak Ground Acceleration} & \multirow{2}{*}{$500 m$} & \multirow{2}{*}{\citet{allen2008}} & $PGA_\mu$ \\ \cline{4-4}
                           &                         &                     & $PGA_\sigma$ \\
\hline
\multirow{2}{*}{Distance to Fault} & \multirow{2}{*}{$30m$} & \multirow{2}{*}{/} & $Dist2F_\mu$ \\ \cline{4-4}
                           &                         &                     & $Dist2F_\sigma$ \\
\hline
\multirow{2}{*}{Distance to Stream} & \multirow{2}{*}{$30m$} & \multirow{2}{*}{/} & $Dist2S_\mu$ \\ \cline{4-4}
                           &                         &                     & $Dist2S_\sigma$ \\
\hline
Geology & 1:50,000 & / & Geo \\
\hline
Bedding & $30m$ & \citet{santangelo2015} & B \\
\hline
\end{tabular}}
\caption{List of covariates used for the modeling phase. The Covariate Name we report corresponds to the covariate we computed at the pixel level, followed by its Resolution or Scale. The Reference describes the method to compute it and the Acronym reports how we will refer to each covariate throughout the text.}
\label{table:CovarList}
\end{table}

Each of the continous covariates listed in Table \ref{table:CovarList} was rescaled with mean zero and unit variance, or in other words, by subtracting every covariate value per SU with the mean covariate value of all the SUs in the study area and ultimately dividing the result by the standard deviation of the covariate values of all the SUs in the area. This procedure ensures that the covariate effects estimated during the modeling phase are all in the same scale, enabling the interpretation of dominant properties on the slope stability process. 

\subsection{Landslide susceptibility via Generalized Additive Model}\label{sec:GAM}

A Generalized Additive Model (GAM) is an extension of the common Generalized Linear Model used in the vast majority of landslide susceptibility studies \citep{reichenbach-2018}. 
The added value corresponds to the ability of estimating both linear and nonlinear relationships between covariates and landslide occurrences. Nonlinearities can be modeled as pure categorical, or more precicely as independent and identically-distributed random variables (\emph{iid}), as well as ordinal, corresponding to a model with adjacent inter-class dependence, which we here implemented as a First Order Random Walk (\emph{RW1}), details provided in \citet{bakka2018,krainski2018}. Such implementation allows one to obtain different regression coefficients for each of the considered portions of a covariate reclassified from a continuous property while simultaneously constraining the ordinal dependence between adjacent classes via a spline interpolation. This procedure has been recently introduced and explained in \citep{Lombardo2018b} in a similar setting although similar modeling approaches have already been tested for landslide susceptibility via machine learning routines such as Multivariate Adaptive Regression Splines \citep[see,][]{conoscenti2016}.

Here we summarize some definitions that will be useful later through the text. A Generalized Linear Model which assumes a Bernoulli probability distribution as the underlying stochastic process can be summarized as follows:

\begin{equation}\label{EQ1}
    \eta(P)=\beta_0 + \beta_1 x_1 + \cdots + \beta_jx_j + f(Xm)
    \end{equation}

where $\eta$ is the logic link, $\beta_0$ is the global intercept, $\beta$ are regression coefficients estimated assuming a linear relationship between landslide occurrence and the given covariate $x_j$. and $f$ can be any function we implement to model discrete covariates ($X_m$). In this work we used a \emph{iid} implementation in case of discrete classes independent from each other, and a \emph{RW1} implementation for ordinal cases, where adjacent classes retain a ordered structure which we want to account for in the modeling stage.  

It is important to note that the right term of Equation \eqref{EQ1} is referred to as the ``linear predictor" and corresponds to the combination of each model component, or in other words, to the sum of all the terms namely, intercept, fixed (covariates linearly modeled) and random effects (covariates nonlinearly modeled).

Once the model estimates the linear predictor, the conversion into probabilities is obtained via the logic link $\eta$ as follows:

\begin{equation}\label{EQ2}
    P= \frac{\beta_0 + \beta_1 x_1 + \cdots + \beta_jx_j + f(Xm)}{1 + \beta_0 + \beta_1 x_1 + \cdots + \beta_jx_j + f(Xm)} 
    \end{equation}

The estimated probabilities can then be intersected with a known (for calibration) and unknown (for validation) vector of presence/absence cases to assess the performance of any model. Here we calibrated over each SU in the area. And, we implemented a 10-fold cross validation scheme where the model is trained with 90\% of the available SUs but constraining the random sampling of the complementary 10\% to extract each SU only once. In turn, the combination of the 10 subsets used for validation returns the whole study area, producing a fully predicted map.

Before moving to the explanation of the statistical simulations, we remind here the reader that any Bayesian model returns a distribution of potential values for each model component, be it the global intercept, each class of a discrete property (both \emph{iid} and \emph{RW1}), the regression coefficients of covariates used linearly and even the outcome (here a probability of landslide occurrence) itself.

\subsection{Statistical Simulations}\label{sec:Simul}

Once a statistical model is estimated, one can always generate any number of simulations from it by randomly sampling from the distribution of each model component and solving for the specific predictive equation. 

This is particularly intuitive in the Bayesian setting where each model component is expressed with a distribution of values. Therefore, a posteriori, one can generate any sequence number of regression coefficient values following the estimated distribution, compute the series of products and sums to calculate the linear predictor and finally transform the result into susceptibilities via the logit link fuction. 
In this work, we use the same structure explained above. However, we apply some changes to retrieve the susceptibility patterns according to past ground shaking, for which we have digital information throughout the history of the Jiuzhaigou area. 

\begin{figure}[t!] 
	\centering
	\includegraphics[width=0.8\linewidth]{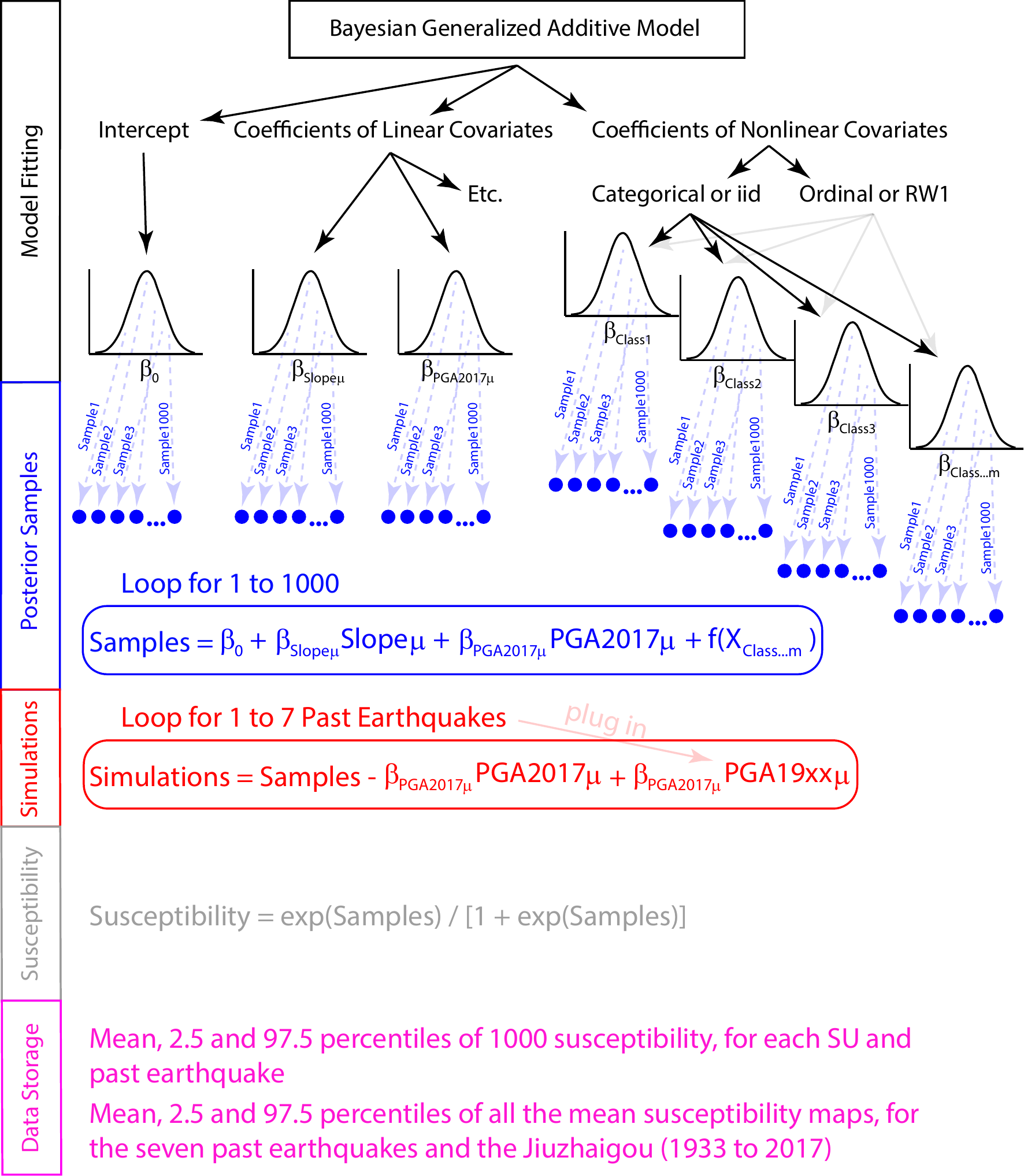}
	\caption{Graphical sketch of the simulation routine.}
	\label{FigureSimul}
\end{figure}

More specifically, we developed an initial model fitted and validated by using co-seismic landslides and the set of covariates listed in Table \ref{table:CovarList}, where the PGA corresponded to the Jiuzhaigou earthquake. Once we tested the prediction skill of the model (reported in Section \ref{sec:Results}) we implemented the following simulation stages, also graphically summarized in Figure \ref{FigureSimul}:

\begin{itemize}
    \item We simulated 1000 realizations of the linear predictor estimated from our binomial GAM implementation for each of the SUs, from the initial model where the ground motion effect onto the susceptibility map was carried by the product of the PGA regression coefficient and the PGA values. Each simulation essential draws one random sample from all regression coefficients' distributions, multiply the sample for the corresponding covariate value and sum all separate model components together. 
    \item From the 1000 simulations we subtracted the product between each randomly generated $\beta_{PGA_{\mu2017}}$ sample and the ${PGA}_{\mu2017}$ values for each SU.  
    \item We added the product between the each randomly generated $\beta_{PGA_{\mu2017}}$ sample and the vector of ${PGA}_{\mu}$ values for each SU, this time coming from one of the 7 past alternatives of PGA pattern listed in Table \ref{table:QuakesList}.
    \item We stored the 1000 simulations, converted them all into susceptibilities by using Equation \eqref{EQ2}.
    \item We calculated the mean susceptibility of the 1000 simulations for each SU and each past earthquake.
    \item We calculated the 95\% Credible Interval (the difference between the 97.5\% and the 2.5\% percentiles of the 1000 simulations) for each SU and each past earthquake.
    \item We calculated the mean and 95\% Credible Interval of all the mean susceptibility maps obtained from 1933 to 2017.
\end{itemize}

\section{Results}\label{sec:Results}

This section includes the assessment of the 10-fold cross-validation routine. This being followed by the description of the significant covariate effects (for which the model is at least 95\% certain that the contribution is either positive or negative, \textit{i.e.,} the estimated distribution does not contain zero). This is complemented by summarizing the reference susceptibility model for the 2017 Jiuzhaigou earthquake in map form and the resulting descripting statistics for the 1000 simulations for each of the 7 earthquake under examination. Ultimately, we show the summary maps for the susceptibility model which combines the signal of all possible susceptibility arrangements for the period between 1933 and 2017. 

\subsection{Predictive performance}\label{sec:Effects}

\begin{figure}[t!] 
	\centering
	\includegraphics[width=\linewidth]{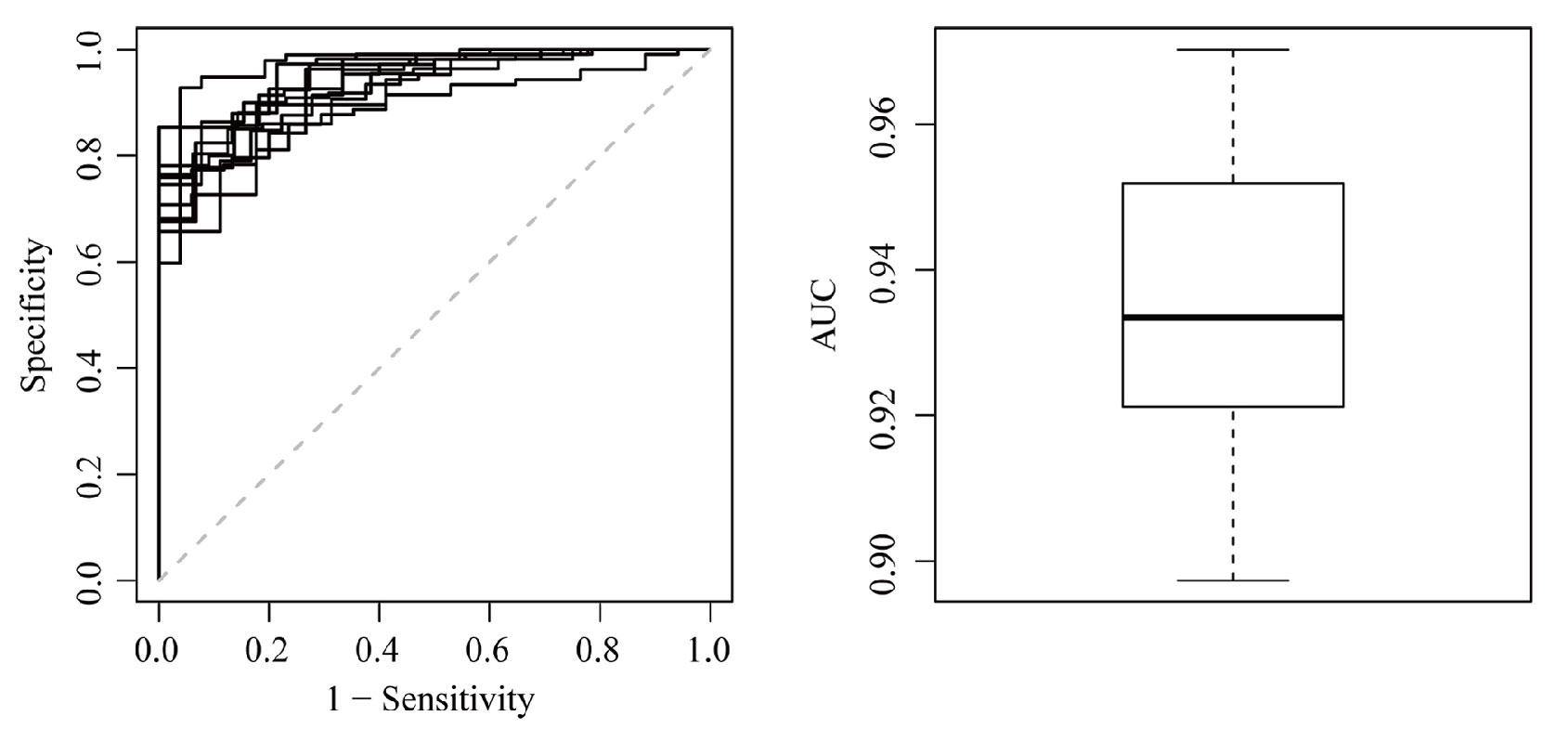}
	\caption{Left panel shows the ten cross-validated Receiver Operating Characteristics curves of the reference model featuring the PGA from the Jiuzhaigou earthquake. The right panel summarizes the associated AUC distribution.}
	\label{Figure6}
\end{figure}

Figure \ref{Figure6} shows the 10-fold cross-validation we performed, which classifies as outstanding according to \citet{hosmer2000}. More specifically, the 10 AUCs we obtained are all confined above 0.9 (with a median of approximately 0.93). And, the variability associated with each cross-validated subset is small. This can be seen both in Figure \ref{Figure6}a where the ROC curves do not spread over the 2D space defined between sensitivity and specificity. The same is also valid for Figure \ref{Figure6}b where the interquartile distance is approximately 0.03 and the difference between minimum and maximum AUC is only 0.07.  

\subsection{Covariates' effects}\label{sec:Effects}

\begin{figure}[t!] 
	\centering
	\includegraphics[width=.55\linewidth]{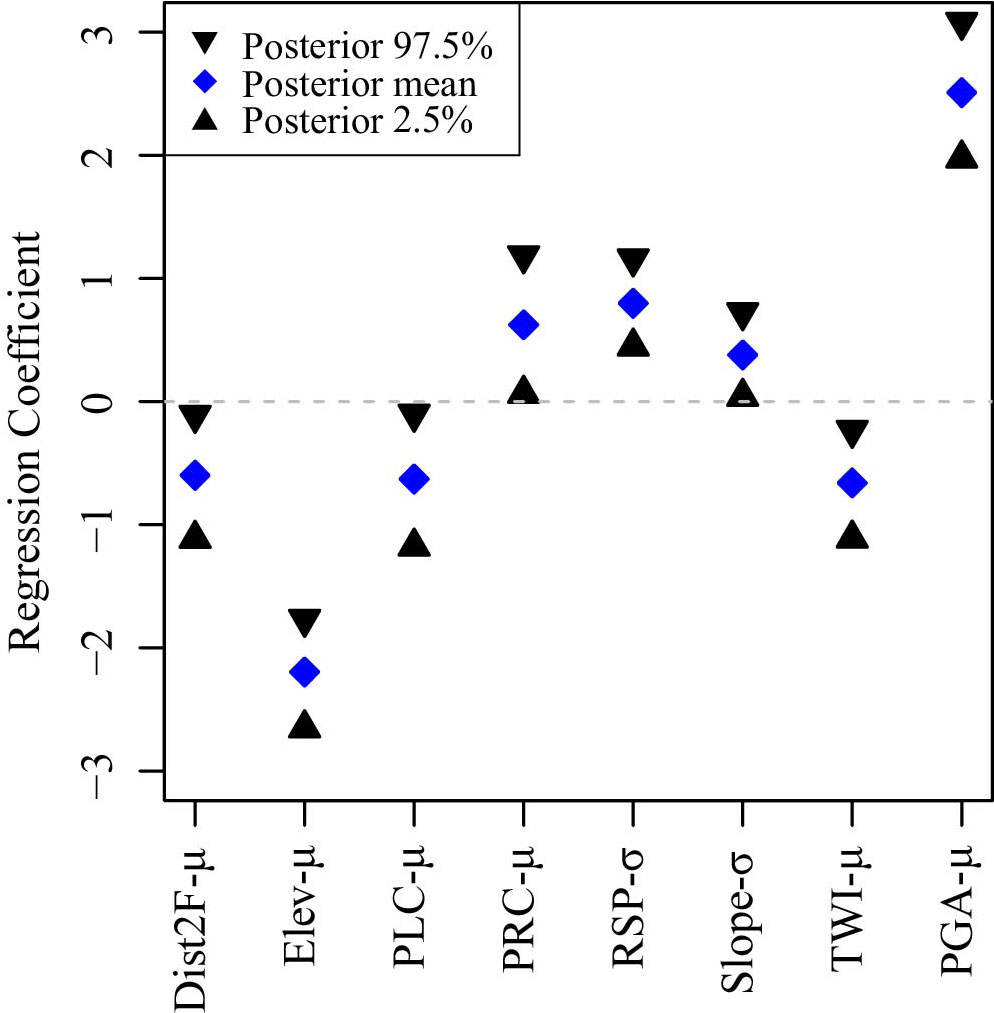}
	\caption{Significant fixed effects shown through their estimated posterior distributions summarized with mean (blue rhombi) and 95\% CI (in black).}
	\label{Figure7}
\end{figure}

Figure \ref{Figure7} shows the significant fixed effects, or the regression coefficients of those covariates that have been used linearly. 

The distribution of all regression coefficients appears quite narrow despite the relatively small sample size, and well determined by the model. The most striking characteristics is associated to the $\beta_{PGA_2017}$ which appears to dominate the susceptibility pattern. In fact, being the covariates rescaled (see, Section \ref{sec:Covar}), the fixed effects reported in Figure \ref{Figure7} are directly comparable, which makes the PGA contribution, in absolute value, much larger than any other covariate, taking aside the $Elev_\mu$, which has an opposite role to the PGA. 

\begin{figure}[t!] 
	\centering
	\includegraphics[width=.8\linewidth]{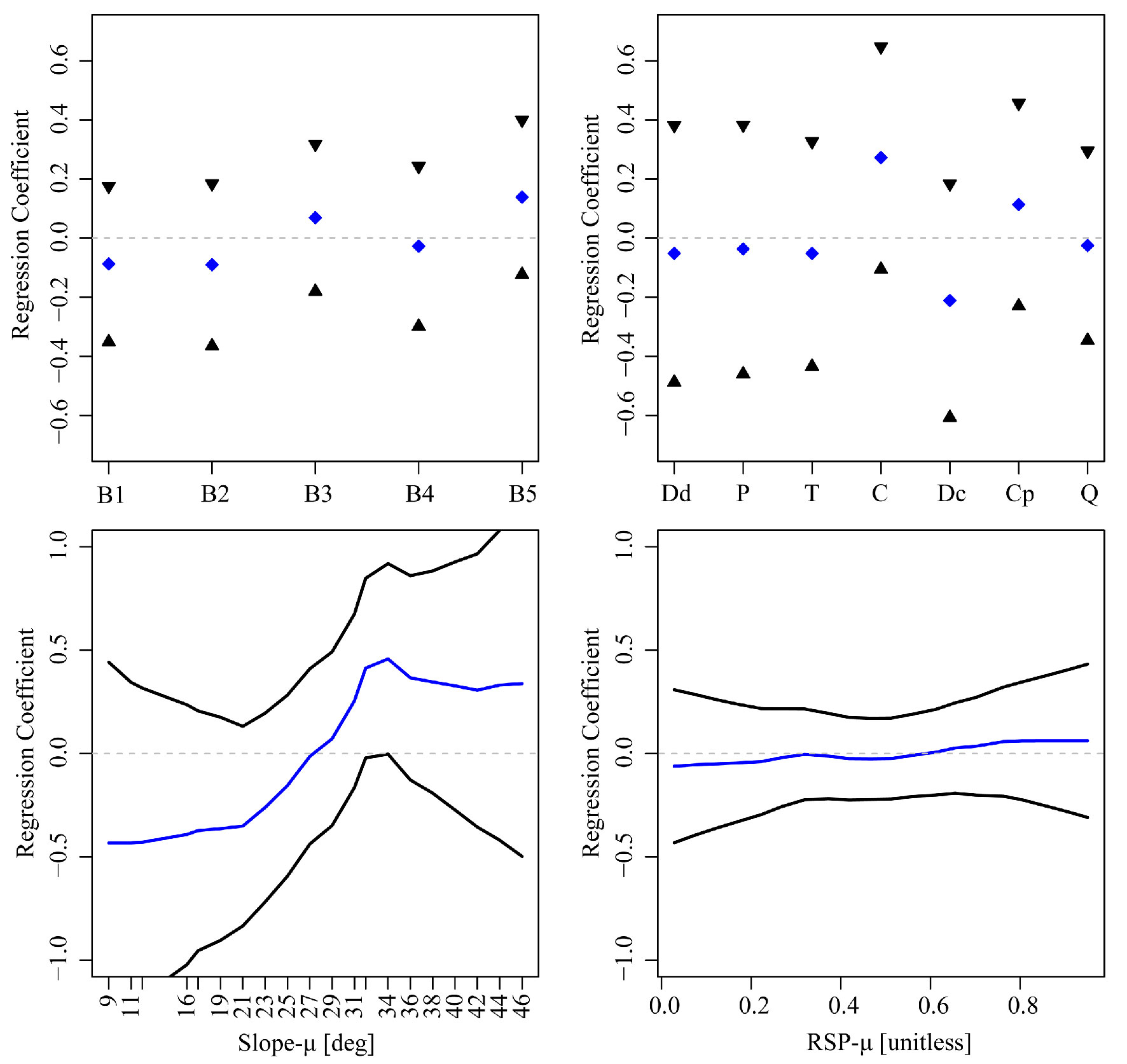}
	\caption{The first row reports the random iid effects, for each class of Bedding and Geology (the acronyms are explained in Tables \ref{table:BeddingTab} and \ref{table:LitoTab}), through their estimated posterior distributions summarized with mean (blue rhombi) the 95\% CI (in black). The second row shows the random ordinal (RW1) effects, of $Slope_\mu$ and $RSP_\mu$ (Relative Slope Position), via their posterior distributions where the mean is highlighted in blue and the the 95\% CI in black.}
	\label{Figure8}
\end{figure}

With regards to the random effects, or those that have been modeled nonlinearly, Figure \ref{Figure8} shows the descriptive statistics of two categorical and two ordinal cases.

Overall, both Bedding and Geology do not appear to be significant (the zero lines cross the distribution of each categorical class) although certain classes have a posterior mean quite far (both positively and negatively) from being null. Therefore, despite the overall non-significance, on average some classes can still contribute to the spatial pattern of the final susceptibility model/map. A similar consideration can be made for $Slope_\mu$ and $RSP_\mu$. Both covariates have their distribution crossed by the zero line. However, the posterior mean of $RSP_\mu$ is clearly aligned with zero making its impact negligible with respect to the final susceptibility model/map. As for the posterior mean of $Slope_\mu$, this is constantly quite far from zero and shows a clear, nonlinear and increasing trend from low to high slope steepness values. More specifically, the $Slope_\mu$ contribution becomes increasingly positive for slopes above 27 degrees of steepness.   

\subsection{Landslide susceptibility mapping}\label{sec:Mapping}

\begin{figure}[t!] 
	\centering
	\includegraphics[width=\linewidth]{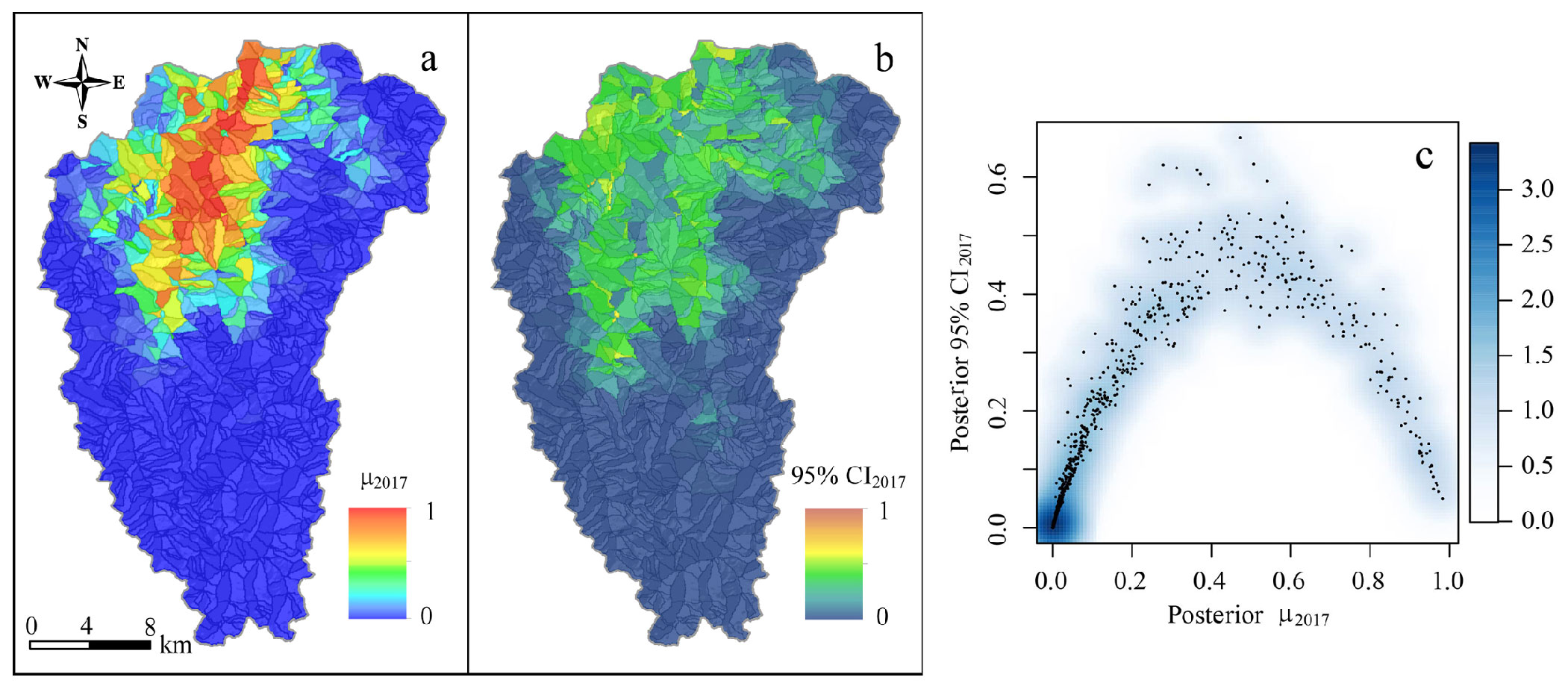}
	\caption{Posterior mean (panel \textbf{a}) and 95\% CI (panel \textbf{b}) of the reference susceptibility model generated with the Peak Ground Acceleration of the Jiuzhaigou earthquake. Panel \textbf{c} shows the error plot where the two maps in the first and second panel are plotted against each other.}
	\label{Figure9}
\end{figure}

Figure \ref{Figure9} shows the summary statistics of the reference model for the 2017 Jiuzhaigou earthquake in map form (posterior mean and 95\% CI), together with the error plot. The mean landslide susceptibility map for the Jiuzhaigou earthquake shows a southward decreasing trend due to the dominant contribution of the $PGA_{\mu_{2017}}$. The uncertainty associated to the mean appears relatively small in the southernmost sector of the study area although it shows a much larger spread to the north. The error plot or mean versus 95\% CI (Figure \ref{Figure9}c) is meant to evaluate whether these estimates are reasonable. In fact, an ideal model should reliably predict very low and very high probability values. In other words, the left and right tail of the posterior mean probability distribution should be associated to a very limited uncertainty. Conversely, the central portion of the posterior mean probability distribution is intrinsically much more difficult (i.e., uncertain) to be determined \citep{reichenbach-2018} and therefore a larger spread can be reasonably accepted \citep{rossi2010}.    

\subsection{Statistical simulations}\label{sec:Simul}

\begin{figure}[t!] 
	\centering
	\includegraphics[width=0.85\linewidth]{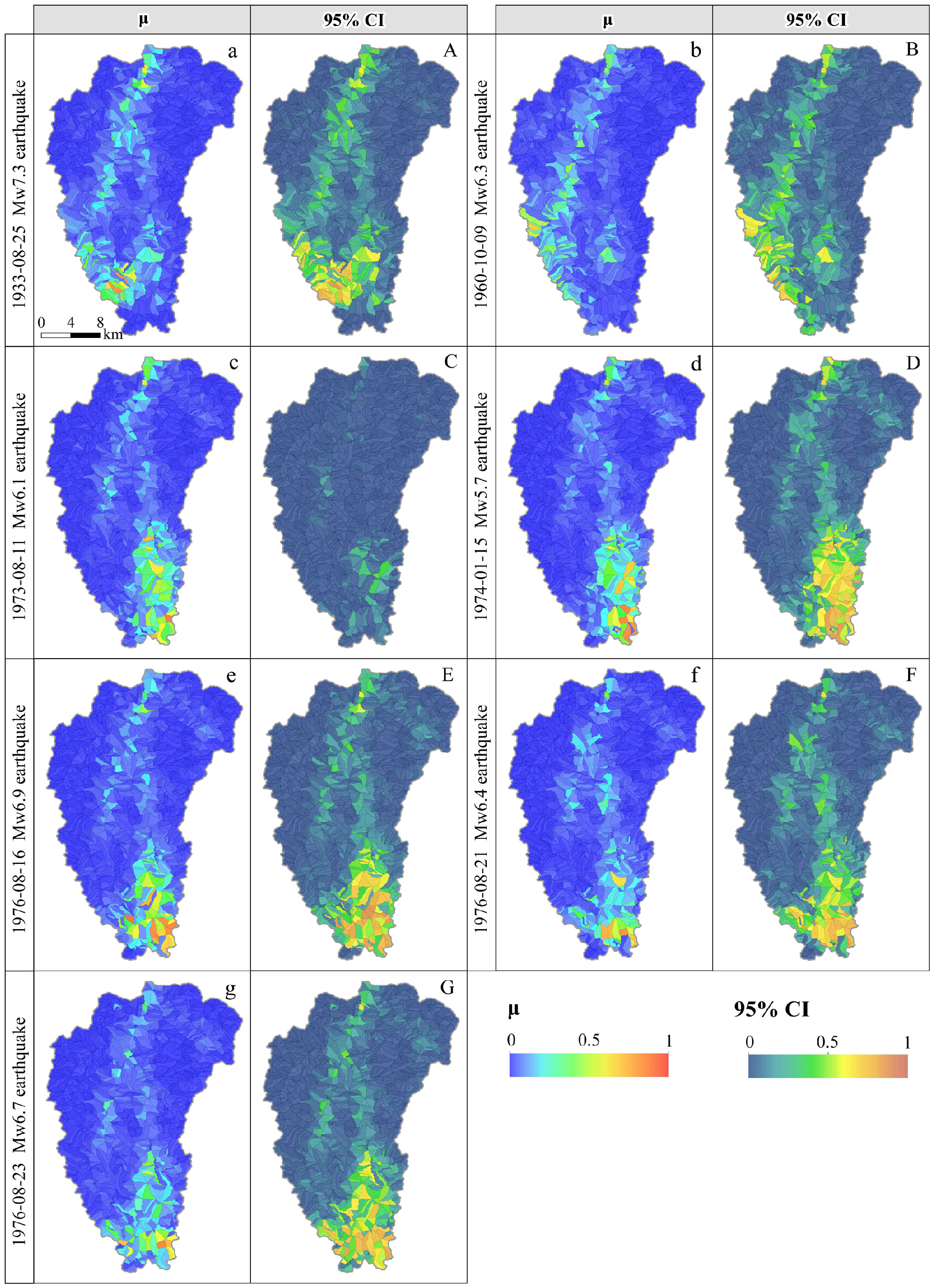}
	\caption{Mean simulated susceptibility maps (a to g) and 95\% CI maps (A to G), for each of the seven earthquakes occurred in the area prior to the Jiuzhaigou earthquake.}
	\label{Figure10}
\end{figure}

For each of the seven other earthquakes (see Table \ref{table:QuakesList}) that occurred in the study area before the Jiuzhaigou earthquake, we also simulated 1000 realizations of the susceptibility patterns by replacing the PGA map of the 2017 Jiuzhaigou earthquake with the ones for the particular earthquakes. To summarize the 1000 simulations we show in Figure \ref{Figure10} the main statistical moment as well as the 95\% CI, separately for each scenario.  

As a result, the susceptibility pattern clearly changes as a function of the various PGA patterns of the earthquake, some of which are located to the south of the study area. More specifically, being the majority of past epicenters located to the south, the largest susceptibility values are shown near the catchment outlet. And, similarly to the spatial patterns shown in Figure \ref{Figure9}, the uncertainty closely follows the susceptibility trend by increasing as the probability increases and decreasing towards the highest mean probability again. This is expected because being the PGA map the largest contributor to the reference model (see Figure \ref{Figure7}), whenever the PGA levels are low, the slopes get estimated with proportionally low susceptibilities. The opposite is also true, for whenever the PGA levels are high, this effect dominates the susceptibility and the other model components have a negligible effect, hence low variations. Conversely, whenever the PGA is in between these two extreme situations, the model becomes more uncertain because the contributions from the other model components becomes more relevant and lead to an increased variability, hence larger uncertainty around the mean susceptibility behavior. 

\begin{figure}[t!] 
	\centering
	\includegraphics[width=\linewidth]{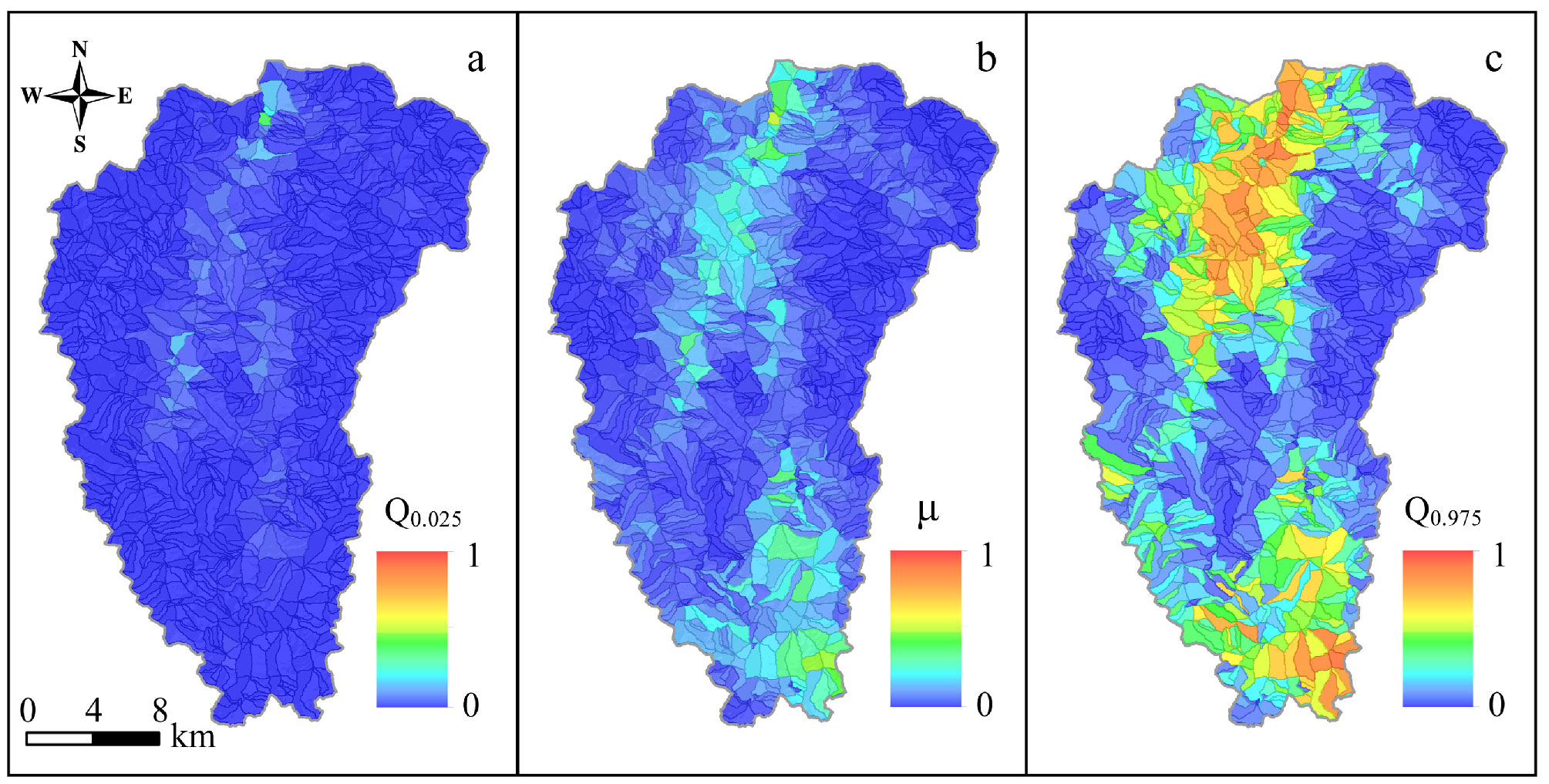}
	\caption{Quantile 0.025 (a), mean (b) and quantile 0.975 (c) of the distribution of all the possible posterior means featuring different ground motion scenarios from 1933 to 2017.}
	\label{Figure11}
\end{figure}

Ultimately, we generated a combined probability which would account for all the possible variations in the susceptibility patterns as a result of the contributions of the ground motions experienced from 1933 to 2017. To achieve this, we combined the mean susceptibility map of the reference model calibrated on the Jiuzhaigou earthquake (Figure \ref{Figure9}) and the mean simulated susceptibility of the other seven earthquakes (Figure \ref{Figure10}). 
Ultimately, to compress the information carried by the multiple scenarios, we compute the mean and 95\% interval across the whole time series. This is shown in Figure \ref{Figure11} where we choose to show the 95\% CI in its separate components, i.e., the 2.5 percentile and the 97.5 percentile to plot the best and worst case scenarios that the area would exhibit across the considered time span. As for most representative pattern since 1933, the mean map delivers such information.   

\section{Discussion}\label{sec:Discussion}

\subsection{Reference model interpretation}\label{sec:RefInterpretation}

In this work, we attempted to combine the ground motion patterns in an overall earthquake-induced landslide susceptibility map for the time period between 1933 and 2017. The reference model which was validated for the Jiuzhaigou earthquake performs with outstanding results \citep{hosmer2000} suggesting that the influence of each model component is well determined (Figure \ref{Figure6}). We prove this in Figure \ref{Figure7} where the regression coefficients can be easily interpreted with respect to the slope instability process, although the same cannot be said for the categorical cases corresponding to the Geology and Bedding. This could be an effect due to the complexity of representing thematic information at the SU level. In fact, one of the most difficult tasks when creating susceptibility models with mapping units different from the grid-cell case consists of the loss of high-resolution information. In fact, at the SU level or catchment or any large mapping unit, one needs to approximate the distribution of properties that vary at small spatial scales. For a continuous factor such as Slope or any other terrain derivative, this comes relatively easy by computing some descriptive statistics such as the mean and standard deviation (same as we did here) or a much finer description into quantiles \citep[e.g.,][]{amato2019}. However, for a geological map or a bedding measurement, representing these two properties at the SU level is more complex. One could opt to assign to a given SU the most representative categorical class (same as we did here) or one could compute the percentage extent of each categorical class with respect to the given SU \citep[e.g.,][]{Camilo2017,Guzzetti2006a}. Here, we opted for the most representative class to minimize the model complexity due to the subsequent simulations stages. However, this may have neglected a more informative use of the Bedding and Geology in the model, which we remind here, outstandingly performed nevertheless.  

With respect to each model component, the mean distance to all the tectonic lines shows a negative contribution (mean $\beta_{Dist2F_{\mu}} = -0.60$). This is generally counter-intuitive as one would expect the closer the seismogenic fault, the higher the probability of landslide occurrence. However, this assumption is not valid for two reasons. The first is that the proximity to the rupturing fault is already part of the PGA information. Therefore, we computed the distance from any fault dissecting the carbonate rocks in the area. As a result, the weakening effect of the fault traces onto the rock mass strength, increases the chance that the loose material draping over the landscape gets removed by regular or common erosion. The regression coefficient of the mean Elevation per SU appears to be negative (mean $\beta_{Elev_{\mu}} = -2.17$), this being a characteristic of the study area, where most of the landslides have all been recognised in the lower portions of the topography (see Figure \ref{Figure2}). It is worth noting that the PGA map, does not account for topographic amplification. Therefore, some confounding effect may still be present between Elevation and PGA. A much more reliable model could actually be obtained by using a ground shaking parameter which includes topographic and possibly soil amplification. By doing this, any landslide predictive model should experience an increase in performance and should provide a much clearer interpretation of each covariate role. 

The mean planar (mean $\beta_{PLC_{\mu}} = -0.63$) and profile (mean $\beta_{PRC_{\mu}} = 0.62$) curvatures show an opposite contribution to the model, where the former favors slope instability in SUs which are predominantly sidewardly concave and the latter contributes to increase the susceptibility in SUs which are upwardly concave. 

The standard deviation of the RSP appears to play a major role in explaining the slope instability. Being the RSP a normalized elevation where the minimum values is assigned to the theoretical floodplain and the maximum value assigned to local ridges, a large standard deviation of this covariate in a given SU implies a large topographic roughness. As a result, the large and positive mean regression coefficient (mean $\beta_{RSP_{\sigma}} = 0.69$) is a reasonable result. A similar signal is carried by the standard deviation of the slope steepness per SU. This covariate is also a proxy for topographic roughness and here (Figure \ref{Figure7}) it is reasonably shown to positively contribute to slope instability (mean $\beta_{Slope_{\sigma}} = 0.38$). The topographic wetness index expresses the morphometric effect to retain water as a function of local slope steepness values and the upslope contributing area. Hence, as the TWI increases it generally expresses locations in the landscape corresponding to flat areas where water flows receiving water flows from upslope, i.e., rivers or floodplains. In this work, we have used the most recent version of the software \emph{r.slopeunits} by \citet{alvioli2016}, which directly removes flat topographies from the SU calculations. And yet, due to the extremely rough landscape, no SUs have been removed. By removing floodplains, one could expect the TWI effect to be positive and to express the effect in pore pressure increase in portion of the slope hanging in the highest sectors of the landscape. However, because \emph{r.slopeunits} could not remove any large flat area, a negative contribution of the TWI may hint for localized conditions near to the river network. This is why we interpret the negative role of the TWI as reasonable in our model (mean $\beta_{TWI_{\mu}} = -0.65$).
Ultimately, the PGA effect is shown to have the largest impact on the susceptibility estimates with a positive contribution (mean $\beta_{PGA_{\mu}} = 2.51$). Because the model recognizes its contribution to be significant and with a narrow credible interval around the mean regression coefficient, we enabled the simulations for previous earthquake occurrences.

Contrary to our expectations, the covariates related to the Lithology (Geology) and structural geology (Bedding) did not appear to be significant. However, this is most probably related to the quality of the input data. Unfortunately the available geological map did not contain information on the detailed lithologies, and contains only a classification based on geological age. More detailed geological mapping would be required to capture the specific lithologies that are more landslide prone. Also the number of measurements for generating the structural geological classes (B1 to B5) might have been too limited. And, the procedure to convert them into meaningful classes by combining the orientation of major discontinuities in relation to the slope steepness and direction, could have been improved, if more data would have been available. Also other relevant factors, such as soil types, and soil depth could not be taken into account due to lack of input data. 

Bedding and Geology did not appear to be significant across each respective classes. Taking the significance aside, some of the classes showed a posterior mean quite different from zero, which implies that the contribution to the posterior susceptibility mean would still be sensitive to such classes. It is worth to note that significance strongly depends on sample size and being the SU 1234 in number, a relatively large credible interval is to be expected. Therefore, here we try to provide an interpretation of the Bedding and Geology roles solely based on the posterior mean contribution, disregarding the rest of the posterior distribution of each class. 

For Bedding, the largest contribution to the the mean susceptibility is carried by B5, or Down-Dip slope (see Figure \ref{Figure12} and Table \ref{table:BeddingTab} in Appendix A), with a mean regression coefficient of 0.24. Despite the limited contribution compared to the other classes in absolute value, B3 (or Reverse oblique slope) also increases the mean susceptibility with a mean regression coefficient of 0.10. This is surprising the most intuitive bedding type would have been B4, also in consideration of the vast majority of landslides being rock-slides. However, the meaning of a nonsignificant covariate indicates that the model is not certain of the covariate role (negative or positive) and therefore, being also the Bedding coefficients close to zero across classes, we can disregard these limited differences and their interpretation with respect to the expected bedding behaviour.      
\begin{figure}[t!] 
	\centering
	\includegraphics[width=\linewidth]{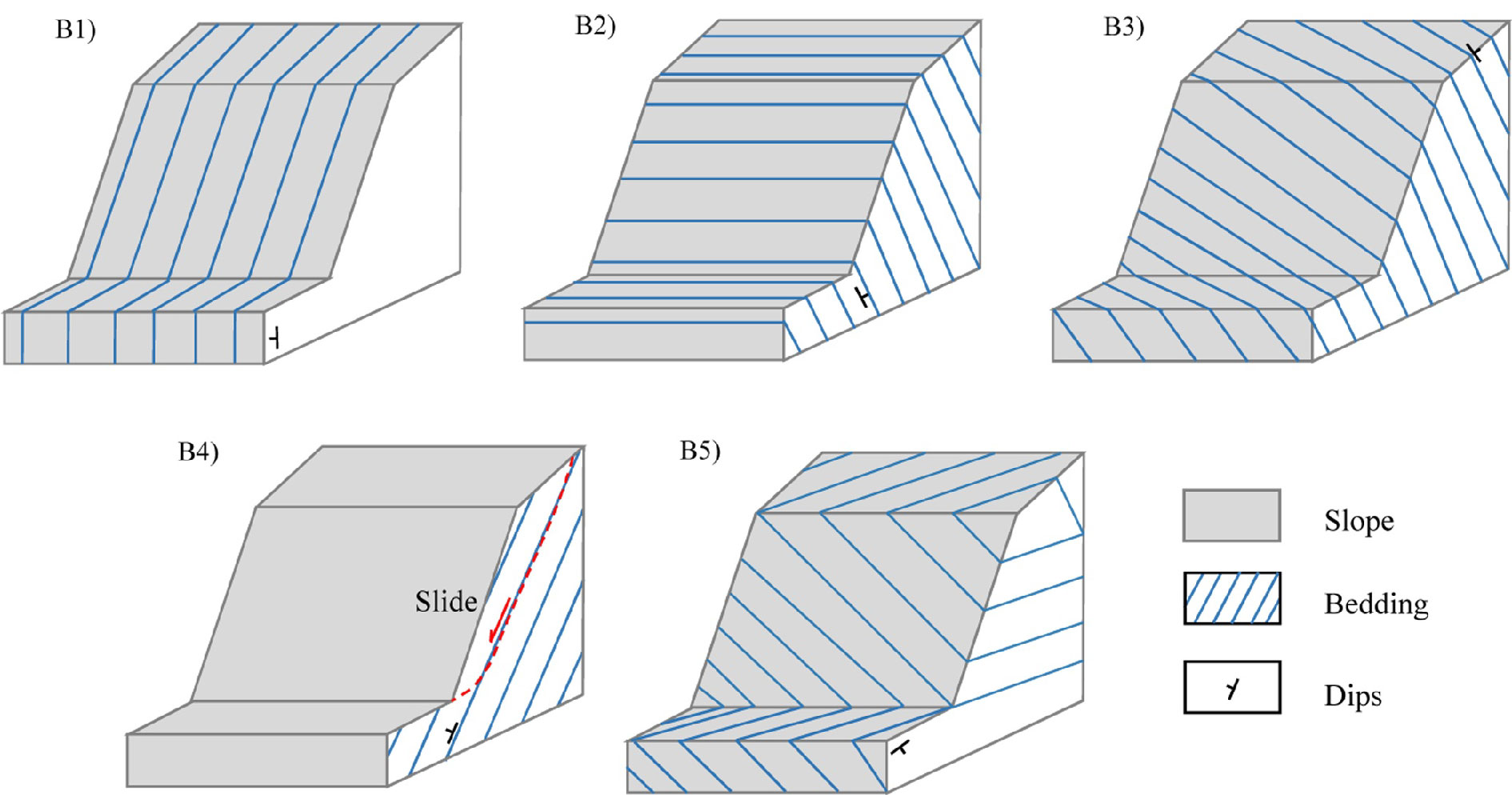}
	\caption{Graphical sketch of Bedding types obtained in the study area.}
	\label{Figure12}
\end{figure}

Similarly, Geology highlights two positive classes, on average, these being C, or Carboniferous limestone, and Cp, or Carboniferous-Permian limestone (see Table \ref{table:LitoTab} in Appendix A), with respective mean coefficients of 0.25 and 0.10. Overall the area essentially comprises calcareous formations whose difference is mainly driven from the fracture system dictated from the tectonic compressive regime. As a consequence, we may infer from a positive mean regression coefficient that lithologies C and Cp may have a higher degree of weathering and fracture network. 

A common test in susceptibility studies to assess how reasonable a model is consists of checking the regression coefficient of the slope steepness. If the slope is estimated to contribute negatively to the model, then either the model is wrong or some variable interaction effects need to be dealt with prior to the modeling phase. Our reference model for the Jiuzhaigou earthquake estimates a positive contribution of the mean slope per SU (see Figure \ref{Figure8}), supporting our assumption that the model reliably recognizes the functional relations between causative factors and landslides. Being the $Slope_\mu$ modeled as a nonlinear ordinal covariate, the posterior mean and 95\% CI trends support this choice. More specifically, the slope classes between 9 and 20 degrees contributes negatively to the landslide susceptibility; and, as of 20 degrees to 34 the contribution to slope instability increases quite linearly becoming positive at 27. From 34 degrees to 46 the contribution does not substantially change.

The $RSP_\mu$ appears to be not significant and even the posterior mean aligns along the zero line (Figure \ref{Figure8}), which in turn means that taking aside the significance, the average contribution of this covariate to the susceptibility is negligible, contrary to what we expected. This position index is a variable that should capture whether landslides are located closer to ridges (high values), mid slope, or lower on the slopes. The fact that there is not so much relation may be caused by the fact that within the slope units, there would be a variation of RSP, and the nature of the slope units may aggregate this variation to an extent where the link between landslides and their triggering location along the topographic profile gets lost.

\subsection{Susceptibility mapping}\label{sec:SuscMapping}

The susceptibility map associated to the Jiuzhaigou earthquake shows a spatial pattern well correlated to the Jiuzhaigou PGA (Figure \ref{Figure9}), implying the dominance of the shaking signal onto the final model. We could only build and validate our model for the Jiuzhaigou case because the only earthquake-induced landslide inventory available in the area corresponds to the co-seismic Jiuzhaigou landslides. Therefore, this model is our reference which we used to infer the PGA effect in the study area over the landslide occurrences and retro-project it to the previous seven earthquakes. 
It is important to note, that we could have modeled the PGA as a nonlinear ordinal property by reclassifying it and using a RW1 same as we did for the $Slope_\mu$ and $RSP_\mu$. However, in doing this we would have calibrated the model on a predefined PGA range, specific of the Jiuzhaigou earthquake (range between 0.08g-0.36g). As a result, it would have been a complex task to extrapolate the PGA effect for the other PGA maps (overall range between 0.02g-0.2g) outside the Jiuzhaigou PGA classified values. For this reason, here we opt to use the Jiuzhaigou $PGA_\mu$ as a linear property, to extrapolate the PGA effect outside the Jiuzhaigou $PGA_\mu$ limit. 
Thanks to this we simulated by using one single parameter distribution for the PGA effect and retrieved one thousand simulated scenarios for each past earthquake (see Figure \ref{Figure10}). As mentioned before, being the past epicenters mostly located to the south of the study area, there the mean simulated susceptibility show the highest values.

This is also reflected in the combined susceptibility maps shown in Figure \ref{Figure11}. The novelty in the simulation procedure we propose is clearly highlighted in this maps which, unless simulated could have not been produced otherwise. In fact, by incorporating different PGA contributions, our combined susceptibility essentially shows the best, average and worst susceptibility scenarios that the study area has theoretically experienced for almost a century. It is important to note that since the other covariates we have used are static (do not significantly change over time), we consider this approach to be valid. However, if other factors such as landuse, roads (cuts and embankments) and buildings (cutslopes) would be incorporated, these would have experienced significant changes in the period since 1933, as the development of the national park has led to many human interventions that might also have contributed to landslide occurrence. Therefore, we suggest that whether simulations in different periods would encompass time-varying covariates, their variation through time should also be expressed and included in the modeling procedure. 

Nevertheless, the combined map we present in Figure \ref{Figure11} is not exactly a conventional susceptibility map as it can be found in many other studies \citep[e.g.,][]{ercanoglu2004,lee2004,westen2008}. Our combined susceptibility incorporates a temporal dimension (limited to the availability of past scenarios) which makes it much closer to the definition of a landslide hazard map. By definition, the landslide hazard should include a return time or the expected frequency of a widespread landsliding event. Here, we propose a map which delivers the slope instability at the SU level for a period of 84 years (1933-2017). However, if a significant earthquake would occur in the direct surroundings of the study area, the landslide pattern might be still quite different, so the map can still not be considered a full predictive map for the coming century.

\section{Conclusions}\label{sec:Conclusions}

Projecting landslide susceptibility maps over temporal scales different from the one responsible for the specific event for which the model is calibrated is quite uncommon in the landslide literature. The very few cases where this is performed correspond to future times, where the land use is expected to change. This is the example of \citet{reichenbach2014} and \citet{pisano2017}, however the implementation they propose neglects the uncertainty that affects the estimation of any regression parameters whereas our implementation follows a greater statistical rigor. 

Our proposed approach is able to depict time-varying susceptibility patterns as a function of the space-time ground motion variability. However, we could only validate the reference model over a specific co-seismic inventory. To test, whether simulations could be reliably made over different temporal scales, more earthquake-induced landslide inventories for the same area should be included to validate the simulations. And, a further improvement could also be made by accounting for additional ground motion effects such as topographic and soil amplifications.

Nevertheless, implementing statistical simulations for earthquake scenarios was never tested so far and especially in the study area, where the main landslide trigger is due to the strong seismicity. Therefore, our proposed method may deliver a much more relevant information to local authorities compared to traditional susceptibility models. In fact, the usual procedure consists of building a susceptibility model trained by using past landslides and either including the responsible ground motion (thus being overly specific) or without it (thus neglecting the spatial dependence in the model induced by the shaking levels).

We also stress here that retrieving past susceptibility patterns are just one application of statistical simulations. In fact, one could also project the simulations towards the future by incorporating scenario-based ground motion. By doing so, one could estimate future landslide-prone areas prior to a potential earthquake occurrence and plan ahead structural design of infrastructure. An example that goes in this research direction can be found in the companion paper to the present one \cite{Lombardo2020}.

\section*{Acknowledgement}
Funding information: this research is financially supported by the National Key R \& D program of China (Grant No. 2017YFC1501002) and Funds for Creative Research Groups of China (Grant No. 41521002). 

\newpage
\appendix
\section{Appendix: Summary of thematic maps}\label{sec:Themes}

\begin{table}[H]
\centering
\begin{tabular}{|c|c|c|}
\hline
\textbf{Bedding acronym} & \textbf{Bedding type} & \textbf{Bedding angle range} \\ 
\hline
B1 & Lateral slope & $60-120$ and $240-300$\\
\hline
B2 & Reverse slope & $150-210$\\
\hline
B3 & Reverse oblique slope & $120-150$ and $210-240$\\
\hline
B4 & Dip slope & $0-30$ and $330-360$\\
\hline
B5 & Down-Dip slope & $30-60$ and $300-330$\\
\hline
\end{tabular}
\caption{Overview of bedding types listed with the same acronyms used throughout the manuscript. The rational corresponds to the relationship we used to calculate the bedding, where \emph{Asp} is the slope aspect [0\degree,360\degree), \emph{Dip} is the dip direction [0\degree,360\degree). From these two values we compute the absolute difference $AbsDiff=|Asp-Dip|$ and reclassify it to obtain the strata attitude or bedding.} 
\label{table:BeddingTab}
\end{table}

\begin{table}[H]
\centering
\scalebox{0.95}{
\begin{tabular}{|c|c|c|}
\hline
\textbf{Geology acronym} & \textbf{Age} & \textbf{Lithotype} \\ 
\hline
Q & Quaternary & Clay, sand and gravel\\
\hline
T & Triassic & Tuffaceous and feldspar/quarz sandstone\\
\hline
P & Permian & Shale, carbonaceous shale with limestone\\
\hline
Cp & Carboniferous/Permian & Brecioid and dolomitic limestone, limestone\\
\hline
C & Carboniferous & Calclithite and compacted limestone\\
\hline
Dc & Devonian & Layered dolomite\\
\hline
Dd & Devonian & Biolithite with argillaceous limestone\\
\hline
\end{tabular}}
\caption{Overview of geological formations where the acronym used throughout the text is associated with the corresponding age and lithotype}. 
\label{table:LitoTab}
\end{table}

\bibliographystyle{CUP}
\bibliography{landslides}
\end{document}